%% file: main.tex
\documentclass[sigconf, 9pt]{acmart}
\usepackage{amsmath,amsfonts}
\usepackage{algorithm}
\usepackage{algorithmic}
\usepackage{graphicx}
\usepackage{textcomp}
\usepackage{xcolor}
\usepackage{subcaption}

\begin{document}

\title{Efficient and Distortion-less Spectrum Multiplexer via Neural Network-based Filter Banks
}
\author{Jiazhao Wang}
    \affiliation{%
      \institution{Singapore University of Technology and Design}
      \country{Singapore}
    }
\author{Wenchao Jiang}
    \affiliation{%
      \institution{Singapore University of Technology and Design}
      \country{Singapore}
    }

\begin{abstract}
Spectrum multiplexer enables simultaneous transmission of multiple narrow-band IoT signals through gateway devices, thereby enhancing overall spectrum utilization. We propose a novel solution based on filter banks that offer increased efficiency and minimal distortion compared with conventional methods. We follow a model-driven approach to integrate the neural networks into the filter bank design by interpreting the neural network models as filter banks. The proposed NN-based filter banks can leverage advanced learning capabilities to achieve distortionless multiplexing and harness hardware acceleration for high efficiency. Then, we evaluate the performance of the spectrum multiplexer implemented by NN-based filter banks for various types of signals and environmental conditions. The results show that it can achieve a low distortion level down to $-39$dB normalized mean squared error. Furthermore, it achieves up to $35$ times execution efficiency gain and $10$dB SNR gain compared with the conventional methods. The field applications show that it can handle both the heterogeneous and homogeneous IoT networks, resulting in high packet reception ratio at the standard receivers up to $98\%$.

\end{abstract}

\maketitle

\pagestyle{plain}
\input{Sections/1_Introduction_New}

\input{Sections/2_Preliminary}
\input{Sections/4_Design_1}
\input{Sections/5_Design_2}

\input{Sections/6_Implementation}
\input{Sections/7_Evaluation}
\input{Sections/8_Discussion}

\bibliographystyle{ACM-Reference-Format}
\clearpage
\bibliography{reference}
\end{document}

%% file: Sections/1_Introduction_New.tex
\section{Introduction}
Spectrum multiplexing is a cornerstone of modern wireless systems, with technologies like 5G-NR~\cite{5gnr} and Wi-Fi 6~\cite{wifi6} utilizing frequency domain multiple access (FDMA)~\cite{goldsmith2005wireless} to serve multiple users. However, these techniques are typically scheme-specific, restricting them to a single communication standard. In the rapidly expanding Internet of Things (IoT) landscape, gateways face the distinct challenge of supporting a multitude of concurrent links from heterogeneous devices, each potentially using a different protocol. This demands a flexible, scheme-agnostic approach to spectrum multiplexing that traditional wireless chipsets, limited by their hardware-centric design, cannot easily provide.

\begin{figure}[!t]
    \centering
    \includegraphics[keepaspectratio=true, width=0.95\linewidth]{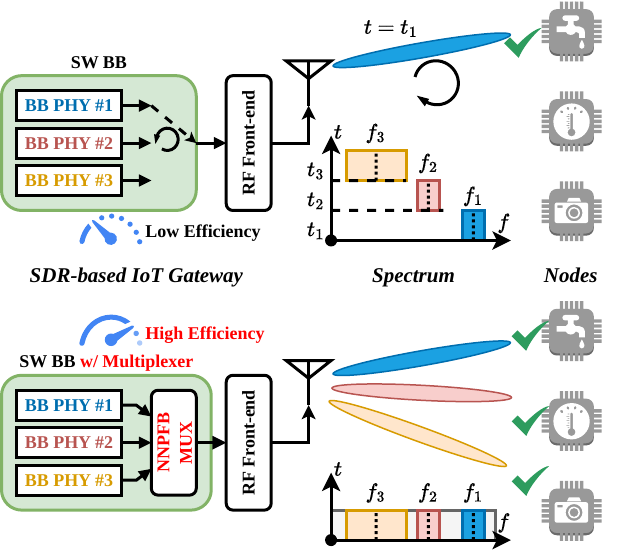}
    \caption{Top: Plain SDR system without spectrum multiplexer module. Bottom: The SDR system with spectrum multiplexer module. }
    \label{fig_SDR_gateway_general}
    \vspace{-5mm}
\end{figure}

\begin{figure*}[t]
    \centering
    \begin{subfigure}[b]{0.4\linewidth}
        \centering
        \includegraphics[keepaspectratio=true,width=\linewidth]{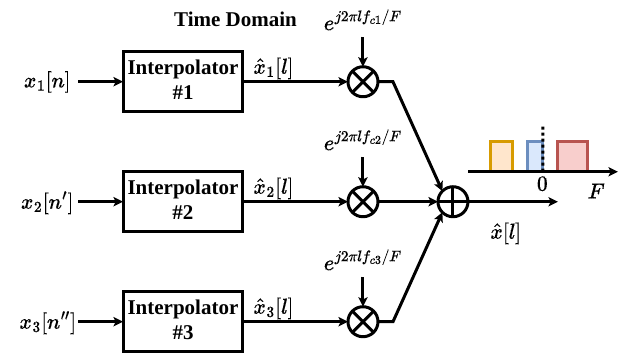}
        \caption{}
        \label{fig_specmux_time}
    \end{subfigure}
    \begin{subfigure}[b]{0.5\linewidth}
        \centering
        \includegraphics[keepaspectratio=true, width=\linewidth]{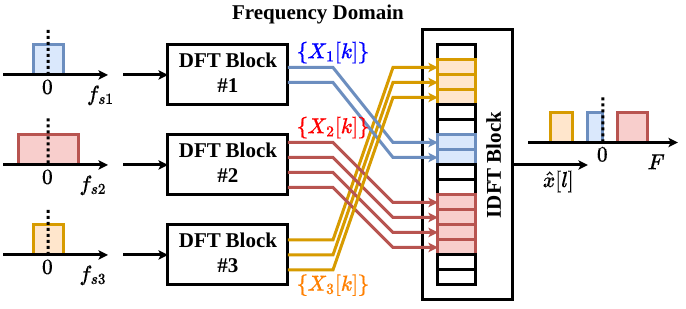}
        \caption{}
        \label{fig_specmux_freq}
    \end{subfigure}
    \vspace{-3mm}
    \caption{(a) Direct spectrum multiplexer based on time-domain interpolation and modulation. (b) DFT-based spectrum multiplexer based on frequency-domain operations.}
\end{figure*}

The software-defined radio (SDR) paradigm (Figure~\ref{fig_SDR_gateway_general}) offers the requisite hardware flexibility to handle diverse communication schemes~\cite{zhang2011wireless, surligas2015empowering}. However, the burden of multiplexing these varied signals efficiently and without distortion falls to the baseband processing algorithms. The challenge, therefore, lies in designing a software multiplexing module that can run on an SDR platform to combine multiple baseband signals into a single wideband signal for transmission. The most straightforward methods for this task are the direct approach, involving time-domain interpolation and modulation~(Figure~\ref{fig_specmux_time}), and the DFT-based approach, which operates in the frequency domain~(Figure~\ref{fig_specmux_freq}). These methods, however, present a fundamental trade-off: the direct approach suffers from high computational latency and requires designing the filter for each baseband stream, while the DFT-based approach introduces significant signal distortion due to spectral leakage~\cite{kistSDRVirtualizationFuture2018}.

To address this trade-off, the \textbf{Oversampled Polyphase Filter Bank~(PFB)} emerges as a theoretically ideal solution. This advanced signal processing structure promises the best of both worlds: the computational efficiency of polyphase implementations and the near-perfect, distortion-free signal reconstruction enabled by oversampling~\cite{harrisDigitalReceiversTransmitters2003, vaidyanathanMultirateDigitalFilters1990}. However, translating this theoretical promise into a practical system is fraught with two significant and distinct challenges. First, the design complexity is formidable; finding the optimal filter coefficients is a highly non-trivial optimization problem that requires deep domain expertise and intricate manual tuning. Second, even with a valid set of coefficients, achieving an efficient implementation that fully leverages the inherent parallelism of the filter bank structure demands specialized software development and careful mapping to hardware architectures.

To overcome these dual barriers of design and implementation, we reframe the entire problem within a machine learning paradigm. We introduce the Neural Network-based Polyphase Filter Bank (NNPFB), where the filter bank's structure is encoded as an interpretable neural network. Our central hypothesis is twofold: (1) that by defining the PFB as a learnable model, we can use powerful gradient-based optimizers to automatically solve the complex coefficient design problem; and (2) that by expressing the solution as a neural network, we can directly leverage the vast ecosystem of highly optimized deep learning libraries and hardware accelerators (e.g., GPUs), thus solving the efficient implementation problem simultaneously. This approach transforms the PFB from a complex, manually-tuned structure into a readily deployable, hardware-accelerated software module.

In summary, this paper makes the following contributions:
\begin{itemize}
\item We address the challenges of oversampled PFB design and implementation by holistically reformulating the problem within an end-to-end learning framework.
\item We propose a novel, interpretable NN architecture that not only automates the discovery of optimal, distortion-free filter coefficients but also inherently maps to highly parallel, hardware-accelerated platforms.
\item We demonstrate through extensive evaluation that our learned NNPFB achieves the signal fidelity of a theoretically ideal oversampled PFB while realizing significant processing gains through hardware acceleration, validating our unified approach to design and implementation.
\end{itemize}

The remaining part of this paper is organized as follows. We first introduce some preliminary on DSP operations in Section~\ref{sec_DSPpre}. Then, we demonstrate how to design the special structures of filter banks for efficient and distortionless spectrum multiplexing in Section~\ref{sec_fbmux}. Next, we transform the filter banks into an interpretable NN-based framework as in Section~\ref{sec_nnmodel}. The implementation and evaluations are in Section~\ref{sec_impl} and \ref{sec_eval}. Finally, we discuss the related works and the potential research directions in Section~\ref{sec_related} and~\ref{sec_disscussion}.

%% file: Sections/2_Preliminary.tex
\section{Preliminary on DSP Operations}\label{sec_DSPpre}
To clarify the upcoming demonstrations, we will first introduce the fundamental DSP operations that are crucial in spectrum multiplexing tasks, including interpolation, decimation, and DFT/IDFT. 

\begin{figure}[h]
    \vspace{-3mm}
    \centering
    \includegraphics[keepaspectratio=true, width=0.9\linewidth]{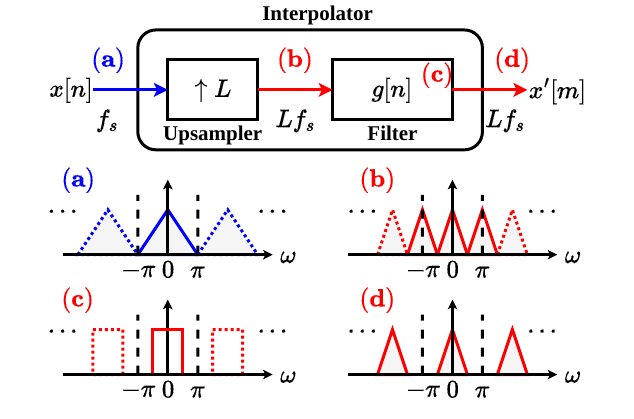}
    \caption{Diagram of interpolators and corresponding spectrum within the process.}
    \label{fig_interpolator}
    \vspace{-3mm}
\end{figure}

\textbf{Interpolation} (Figure~\ref{fig_interpolator}) is a process used to increase the sampling rate of an input signal sequence. It consists of two main components: an upsampler and an anti-imaging filter. The original discrete signal sequence $x[n]$ has a periodic spectrum as in Spectrum~\textbf{(a)}. Here, $\omega$ represents the normalized frequency, defined as $\omega=2\pi \frac{\text{analog frequency}}{\text{sample frequency}}$. The $1$-to-$L$ upsampler increases the sampling rate from $f_s$ to $Lf_s$ by inserting $L-1$ zeros between the original samples. However, the upsampling introduces image components in the normalized frequency range $[-\pi,\pi]$, shown in Spectrum~\textbf{(b)}. To eliminate these unwanted spectral components, the upsampled sequence is passed through an anti-imaging filter $g[n]$, which ideally has a rectangular frequency response. The formulation for interpolation is provided in Equation~\ref{eq_interp}, where $f(m)$ represents the anti-imaging filter. 

\begin{equation}
    x'(m) = \sum_{n=-\infty}^{+\infty}g(m-rL)x(n)
    \label{eq_interp}
\end{equation}

\begin{figure}[h]
    \vspace{-3mm}
    \centering
    \includegraphics[keepaspectratio=true, width=0.9\linewidth]{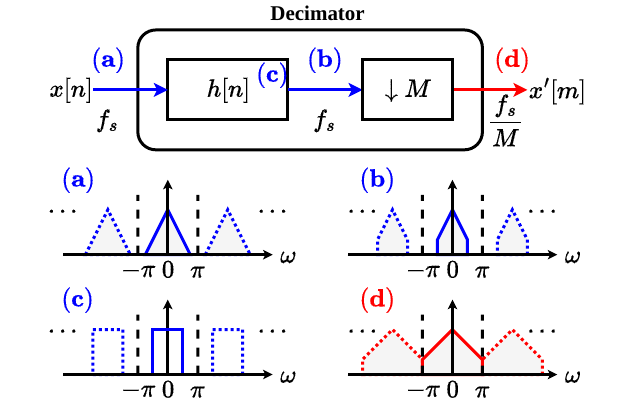}
    \caption{Diagram of decimators and corresponding spectrum within the process.}
    \label{fig_decimator}
    \vspace{-3mm}
\end{figure}

\textbf{Decimation} (Figure~\ref{fig_decimator}) is to reduce the sampling rate of an input signal sequence. It involves two key components: an anti-aliasing filter and a downsampler. The downsampler, which keeps one sample from every $M$ samples, can cause spectral expansion, leading to aliasing. To prevent this, the signal is first filtered with an anti-aliasing filter. The decimation process is described in Equation~\ref{eq_decim}. 

\begin{equation}
    \begin{aligned}
        x'(m) &= \sum_{k=-\infty}^{+\infty}h(k)x(Mm-k) \\
            &= \sum_{n=-\infty}^{+\infty}h(Mm-n)x(n)
    \end{aligned}
    \label{eq_decim}
\end{equation}

\textbf{DFT/IDFT} are fundamental blocks in digital signal processing, primarily used for converting between time-domain signal samples and frequency-domain spectral components. The discrete Fourier transform (DFT) transforms a sequence of $N$ complex samples, ${\mathbf{x}_n} = (x[0], x[1], \dots, x[N-1])$, into a corresponding sequence of complex frequency components, ${\mathbf{X}_k} = (X[0], X[1], \dots, X[N-1])$. Conversely, the inverse discrete Fourier transform (IDFT) converts the frequency-domain sequence ${\mathbf{X}_k}$ back to the time-domain sequence ${\mathbf{x}_n}$. This process is described in Equation~\ref{eq_dft_idft}.

\begin{equation} 
    \begin{aligned}
        DFT: X[k] &= \frac{1}{N}\sum_{n=0}^{N-1}x[n]W_K^{-kn} \\
        IDFT: x[n] &= \frac{1}{N}\sum_{k=0}^{N-1}X[k]W_K^{kn} \\
        W_K = e^{j\frac{2\pi}{K}}
    \end{aligned}
    \label{eq_dft_idft}
\end{equation}

%% file: Sections/4_Design_1.tex
\section{Filter-Bank-based Spectrum Multiplexer}\label{sec_fbmux}
We start with outlining the basic architecture of filter banks, followed by a demonstration of how they can be adapted for spectrum multiplexing, including the structural adjustments and considerations for filter design.

\subsection{Basic Architecture}
\begin{figure}[t]
    \centering
    \begin{subfigure}[b]{\linewidth}
        \centering
        \includegraphics[keepaspectratio=true,width=\linewidth]{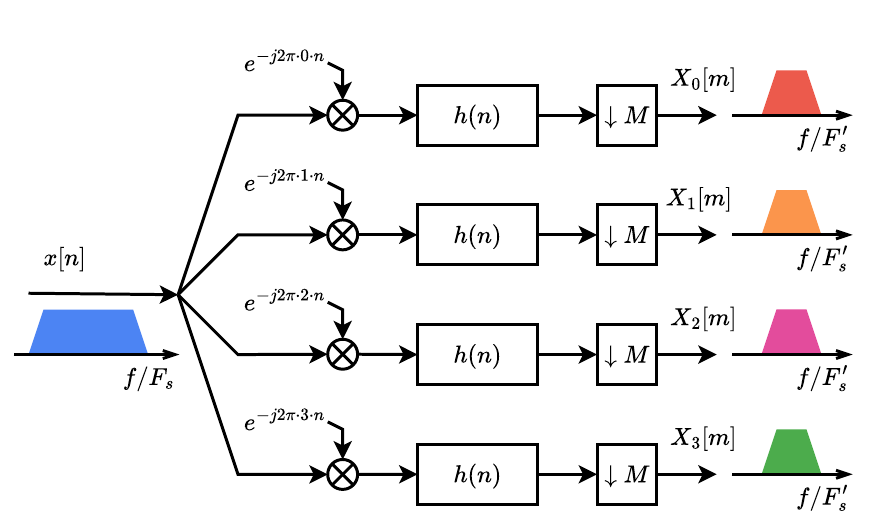}
        \caption{Analysis filter bank}
        \label{fig_AFB}
    \end{subfigure}
    \begin{subfigure}[b]{\linewidth}
        \centering
        \includegraphics[keepaspectratio=true,width=\linewidth]{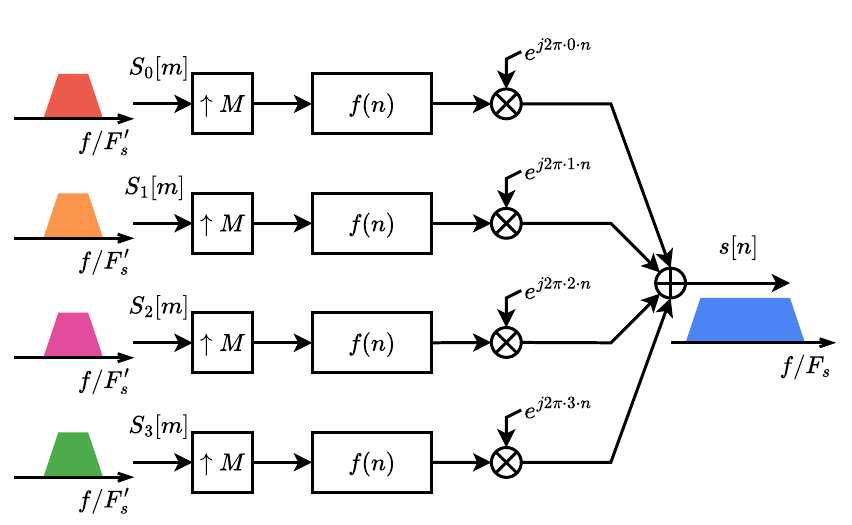}
        \caption{Synthesis filter bank}
        \label{fig_SFB}
    \end{subfigure}
    \vspace{-5mm}
    \caption{ Diagrams of (a) the analysis filter bank with $4$ arms, (b) the synthesis filter bank with $4$ arms.}
    \label{fig_filterbanks}
    \vspace{-5mm}
\end{figure}

The core principle of the filter-bank-based spectrum multiplexer is the decomposition and recombination of signals. The analysis filter bank decomposes an input baseband signal into multiple sub-band signals, each with a reduced bandwidth and a lower sample rate. These sub-band signals are then recombined by the synthesis filter bank to create a signal stream with a larger bandwidth and higher sample rate. The typical structures of AFBs and SFBs are shown in Figure~\ref{fig_AFB} and \ref{fig_SFB}.

In the AFB example in Figure~\ref{fig_AFB}, the input signal $x[n]$ represents baseband signal samples generated at a sample rate of $B$ Hz, thus occupying a bandwidth of $B$ Hz. The objective is to decompose $x[n]$ into $4$ sub-bands, each representing a portion of the original signal centered at the normalized frequency $f_{sub}^{(k)} = k/4$ for the $k$-th branch ($k = 0, 1, 2, 3$). To extract these sub-band signals, we shifted the signal to zero frequency by modulating $x[n]$ with $4$ corresponding complex exponential sequences and then filtered the modulated signals with a low-pass filter $h[n]$ with a bandwidth of $B/4$ Hz. The filtered sub-band signals retain the original sample rate of $B$ Hz but have a reduced bandwidth of $B/4$ Hz. These signals can then be downsampled by $M$ times. The resulting sub-band signals, $X[m, k]$, represent the time-frequency domain components of the original signal, as described in Equation~\ref{eq_afb}, where $x(n)W_K^{-kn}$ denotes the modulated input. This signal is filtered using the filter $h[n]$ and then downsampled by $M$ times, as detailed in Section~\ref{sec_DSPpre}.

\begin{equation}
    X(m,k) = \sum_{n=-\infty}^{+\infty} h(mM-n)x(n)W_K^{-kn}
    \label{eq_afb}
\end{equation}

The synthesis filter bank (SFB) performs the reverse process, as shown in Figure~\ref{fig_SFB}. The input signals, $S[m, k]$, are sub-band signals with the same sample rate and bandwidth. To generate a signal with a wider synthesized bandwidth, these sub-band signals are upsampled to a higher sample rate. Since all sub-band signals have identical sample rates and bandwidths, the same interpolation configuration can be applied across the sub-band signals. Each sub-band signal is upsampled by $L$ times and filtered using an anti-imaging filter $f[n]$. The upsampled signals are then shifted to their designated frequency channels through dedicated modulators. The entire SFB process is described in Equation~\ref{eq_sfb}.

\begin{equation}
    s(n) = \sum_{k=0}^{K-1}W_K^{kn}\sum_{m=-\infty}^{+\infty}S(m, k)f(n-Lm)
    \label{eq_sfb}
\end{equation}

\subsection{Oversampled Filter Banks}
\begin{figure}[h]
    \vspace{-3mm}
    \centering
    \includegraphics[keepaspectratio=true, width=0.8\linewidth]{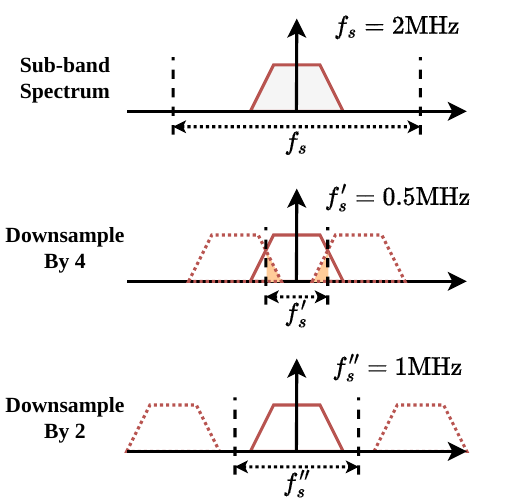}
    \vspace{-2mm}
    \caption{The spectrum of sub-band signal and downsampled versions. The dashed components represent the periodic replicates of discrete-time signals.}
    \label{fig_spec_os}
    \vspace{-5mm}
\end{figure}

Power leakage in AFBs caused by the imperfect filter response~\cite{harrisDigitalReceiversTransmitters2003, harrisWideband160channelPolyphase2011, harrisCascadeNonmaximallyDecimated2017} can lead to significant aliasing when the sub-band signals are downsampled by a factor equal to the number of total sub-bands. For instance, as shown in Figure~\ref{fig_spec_os}, when the signal extracted from sub-band $0$ is downsampled by a factor of $4$, spectral overlap (indicated by the orange shadow in the middle figure) becomes noticeable. Here, the sample rate of each sub-band signal is equal to the frequency interval between sub-bands.

To address this issue, we draw inspiration from the oversample filter banks introduced in \cite{crochiereMultirateDigitalSignal1983}, which help mitigate power leakage and improve performance. \textit{Oversample} refers to the case where the sampling rate of the sub-band signals is higher than the sub-band frequency interval. The downsampling factor is set to be smaller than the number of sub-bands, typically defined as $K = MI$, where $K$ is the number of sub-bands, $M$ is the downsampling factor, and $I$ represents the oversampling ratio. For example, as shown in Figure~\ref{fig_spec_os}, we set $M = 2$. As a result, no spectral aliasing occurs in the downsampled sub-band signal.

We also implement an over-sample synthesis filter bank using a similar configuration, where $K = LI$, and the over-sample ratio $I$ matches that used on the AFB side.





\subsection{Stateless Polyphase Decomposition}\label{subsec_ppfb}
We further apply polyphase decomposition to the filter banks, which can significantly reduce the redundancy within the interpolation/decimation and modulation. 

Although researchers \cite{harrisDigitalReceiversTransmitters2003, harrisWideband160channelPolyphase2011, harrisCascadeNonmaximallyDecimated2017} proposed an efficient implementation of polyphase decomposition of over-sample filter banks on FPGAs, such the method requires a finite state machine to alternate the input indexing depending on the output indices, which is not friendly for our NN-based design. Instead of the \textit{stateful} design, we intend to apply a \textbf{stateless} polyphase decomposition strategy with only simple index rearranging, making it more convenient to transform into NN models.

As for AFBs, given the $k$-th sub-band signal from total $K$ branches, we make the change of variables with $n=rK + \rho, \quad \rho = 0, 1, ... , K - 1$, so the $k$-th sub-band signal can be expressed as
\begin{equation}
    \begin{aligned}
        X(m, k) &= \sum_{r=-\infty}^{+\infty}\sum_{\rho=0}^{K-1} h(mM-rK-\rho) x(rK+\rho)W_K^{-k\rho} \\
        &= \sum_{\rho=0}^{K-1}W_K^{-k\rho}  \left[ \sum_{r=-\infty}^{+\infty}p_{\rho}(m-rI)x_{\rho}(r) \right]
    \end{aligned}
    \label{eq_ppafb_1}
\end{equation}

Equation~\ref{eq_ppafb_1} illustrates the stateless polyphase decomposition strategy. The input signal $x(n)$ is decomposed into $K$ branches, and we denote them as $x_{\rho}(r)=x(rK+\rho)$. Similarly, we denote $p_\rho(m)=h(mM-\rho)$ for the filter decomposition.

By substituting $K = MI$ into the equation, it becomes evident that the term in brackets in Equation~\ref{eq_ppafb_1} defines an interpolator with an upsampling factor of $I$. The term $\sum_{\rho=0}^{K-1}W_K^{-k\rho}$ represents a discrete Fourier transform (DFT) process applied after interpolation.

Similarly, the structure of the over-sampled polyphase synthesis filter for the case $K = LI$ can be derived in a similar manner. After making the change of variables with $n=rK+\rho, \quad \rho = 0, 1, ..., K - 1$,  and exchanging the order of summation, we can get the synthesized signal as
\begin{equation}
    s(rK + \rho) = \sum_{m=-\infty}^{\infty}f(rK+\rho-mL)\left[ \sum_{k=0}^{K-1}S(m,k)W_k^{k\rho}\right]
    \label{eq_ppsfb_1}
\end{equation}

Then, we denote the decomposition of the output signal $s(n)$ and the filter $f(n)$ as
\begin{equation}
    \begin{aligned}
        s_{\rho}(r) &= s(rK + \rho) \\
        q_{\rho}(m) &= f(mL+\rho)
    \end{aligned}
\end{equation}

and we can rewrite Equation~\ref{eq_ppsfb_1} as
\begin{equation}
    \begin{aligned}
        s_{\rho}(r) &= \sum_{m=-\infty}^{\infty} q_{\rho}(rI-m) \left[ \sum_{k=0}^{K-1}S(m,k)W_k^{k\rho}\right] \\
                    &= \sum_{m=-\infty}^{\infty} q_{\rho}(rI-m) \hat{S}_{\rho}(m)
    \end{aligned}
    \label{eq_ppsfb_2}
\end{equation}

\begin{figure}[t]
     \centering
     \begin{subfigure}[b]{\linewidth}
         \centering
         \includegraphics[keepaspectratio=true,width=\linewidth]{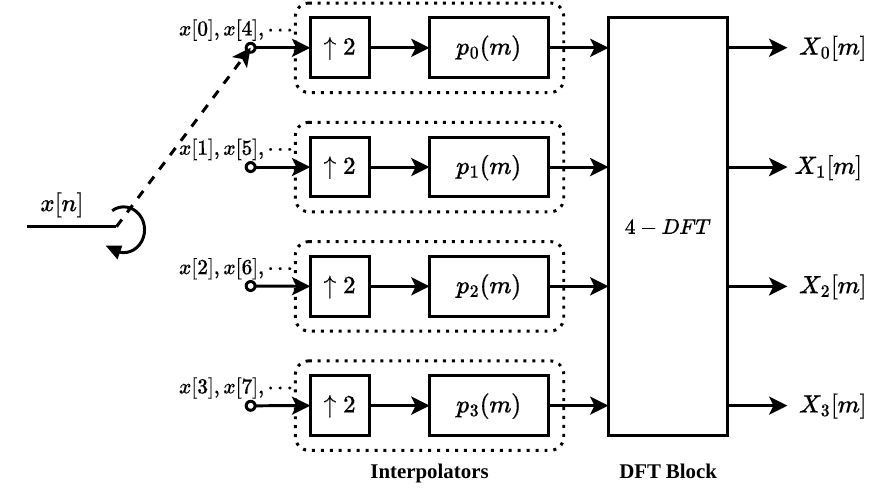}
         \caption{Over-sampled polyphase analysis filter bank}
         \label{fig_PAFB}
     \end{subfigure}
     \begin{subfigure}[b]{\linewidth}
         \centering
         \includegraphics[keepaspectratio=true,width=\linewidth]{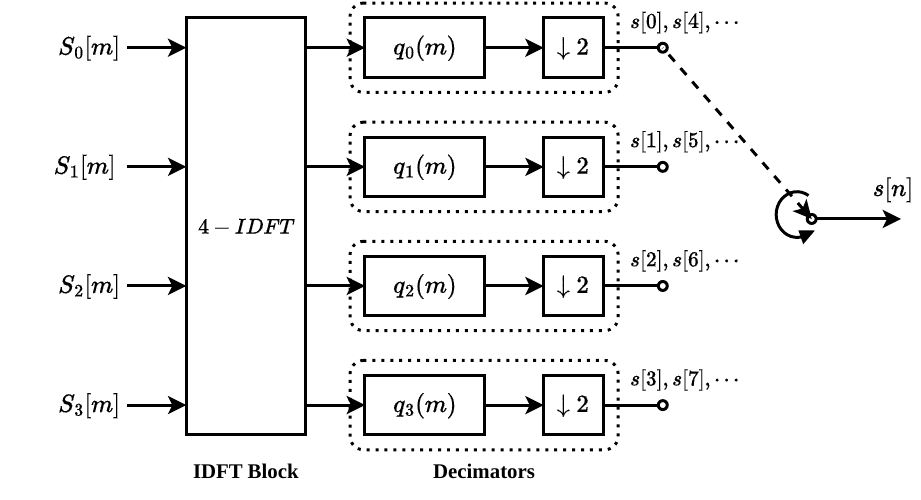}
         \caption{Over-sampled polyphase synthesis filter bank}
         \label{fig_PSFB}
     \end{subfigure}
     \vspace{-2mm}
     \caption{Diagrams of (a) the analysis filter bank with $4$ arms and (b) the synthesis filter bank with $4$ arms.}
     \label{fig_OSFB}
     \vspace{-5mm}
\end{figure}

The term, $\hat{S}_{\rho}(m) = \sum_{k=0}^{K-1}S_k(m)W_k^{k\rho}$, represents the inverse discrete Fourier transform (IDFT) of $S[m,k]$. This expression takes a form similar to decimation with a downsampling factor of $I$ in each branch. Based on this, we can derive the structure of over-sampled polyphase filter banks, as shown in Figure~\ref{fig_OSFB}.

Therefore, we can derive the pipeline of over-sample polyphase filter banks with the structure in Figure~\ref{fig_OSFB}. We don't need any stateful operation for input/output, which indicates the stateless polyphase decomposition. Compared with the basic structure, the interpolation and decimation on each branch are much less complex, as the decomposed filters are much shorter. Furthermore, the modulation operations on each branch are also simplified with fixed DFT/IDFT operations. It is important to note that the DFT/IDFT operations used in AFB/SFB do not signify a transformation between the time domain and the frequency domain. Instead, they function as computational patterns within the filter bank structure.

%% file: Sections/5_Design_2.tex
\section{NN-based Spectrum Multiplexer}\label{sec_nnmodel}
Given the pipeline of the filter banks, we transform the filter banks into neural network models using a model-driven approach. This transformation can take two key advantages of neural networks: first, their learning ability aids in filter weight derivation; second, robust and general support for hardware acceleration, which greatly enhances processing efficiency.

\subsection{NN as DSP Blocks}
The key insights that drive us to apply NN as DSP blocks are

\begin{itemize}
    \item \textbf{Observation 1:} The computation mechanisms of several neural network layers share similar patterns to the DSP operations, so we derive the direct connections between the configuration of the neural network layers and the DSP blocks.

    \item \textbf{Observation 2:} The complex values in DSP operations can be represented by $2$D vectors of corresponding real and imaginary parts, so we fit the complex-valued operations into real-valued neural network layers. 
\end{itemize}

To better demonstrate these, we discuss the typical computation mechanism of transposed convolutional and convolutional layers. 

\begin{figure}[t]
    \centering
    \begin{subfigure}[b]{\linewidth}
        \centering
        \includegraphics[width=0.9\linewidth]{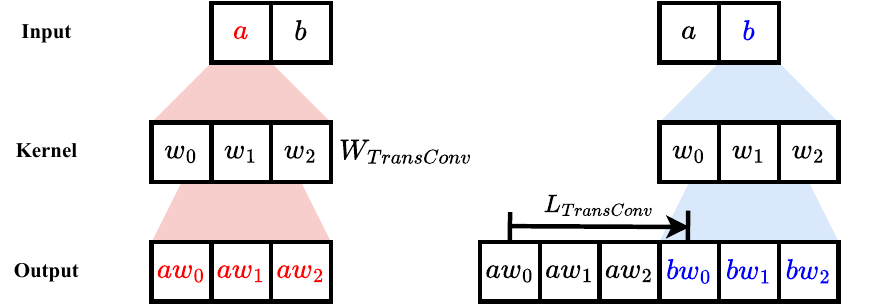}
        \caption{}
        \label{fig_basicTransConv}
    \end{subfigure}
    \begin{subfigure}[b]{\linewidth}
        \centering
        \includegraphics[width=0.9\linewidth]{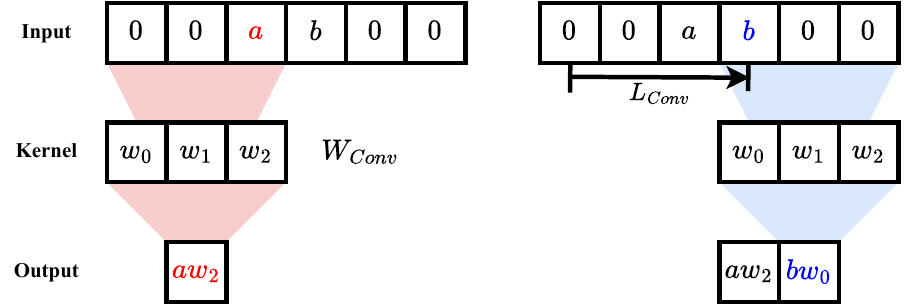}
        \caption{}
        \label{fig_basicConv}
    \end{subfigure}
    \caption{Diagram of the basic operation of (a) the transposed convolutional layer and (b) the convolutional layer.}
    \vspace{-2mm} 
\end{figure}

A transposed convolutional layer is a type of neural network layer that applies a sliding kernel to input data, producing an output with a larger spatial dimension. The $1$-D transposed convolutional layer with single input and output channel is illustrated in Figure~\ref{fig_basicTransConv}. 

The elements in the input $[a,b]$ are multiplied by a \textit{kernel}, $W_{TransConv}$. The multiplication results are mapped to the output successively. The step between each multiplication result is determined by the \textit{stride}, $L_{TransConv}$ parameter. The output is generated by summing the results of the multiplications. Therefore, the computation model of the transposed convolutional layer is similar to the interpolation process (Equation~\ref{eq_interp}) as  Equation~\ref{eq_transConv}.

\begin{equation}
    y[n] = \sum_{k=0}W_{TransConv}[n-kL_{TransConv}]x[k]
    \label{eq_transConv}
\end{equation}

As for the convolutional layer, it applies a sliding kernel to input data, producing an output with a smaller spatial dimension as shown in Figure~\ref{fig_basicConv}.

Similar to the transposed convolutional layer, the $1$-D convolutional layer applies a sliding kernel, $W_{Conv}$, to the input. Usually, $W_{Conv}-1$ zeros are padded on each side of the input sequence. The multiplication results are mapped to the output. The step of kernel sliding is determined by the \textit{stride}, $L_{Conv}$ parameter. The process can be modeled as follows, which is similar to the decimation process in Equation~\ref{eq_decim} but differs in the indexing. 
\begin{equation}
    y[n] = \sum_{k}W_{Conv}[k]x[nL_{Conv}+k]
    \label{eq_conv}
\end{equation}

\begin{figure}[t]
\centering
\begin{subfigure}[b]{0.8\linewidth}
    \centering
    \includegraphics[keepaspectratio=true,width=\linewidth]{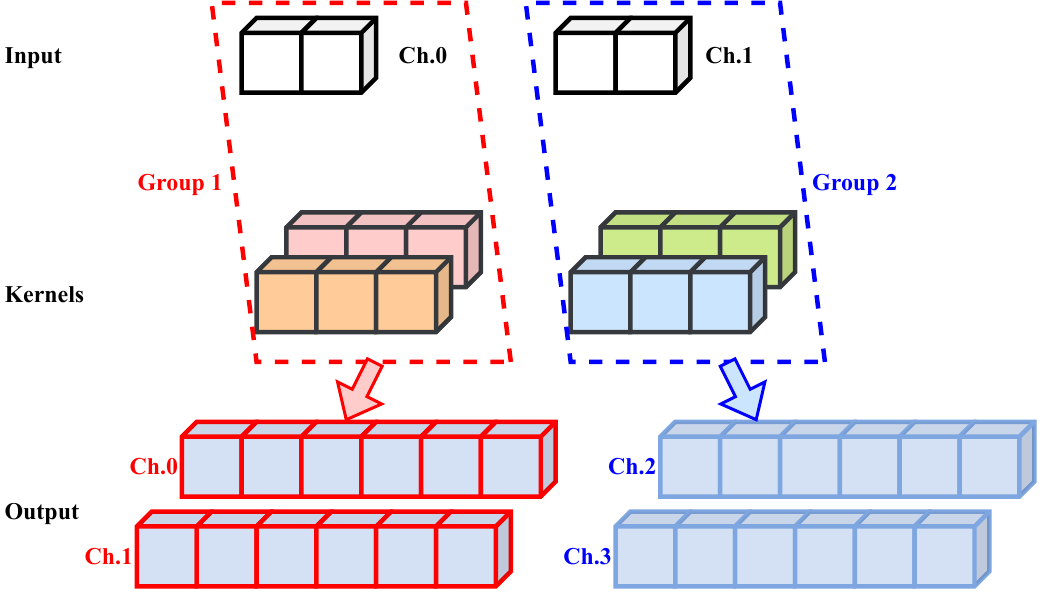}
    \caption{}
    \label{fig_multiChanConvTrans}
\end{subfigure}
\begin{subfigure}[b]{0.8\linewidth}
    \centering
    \includegraphics[keepaspectratio=true,width=\linewidth]{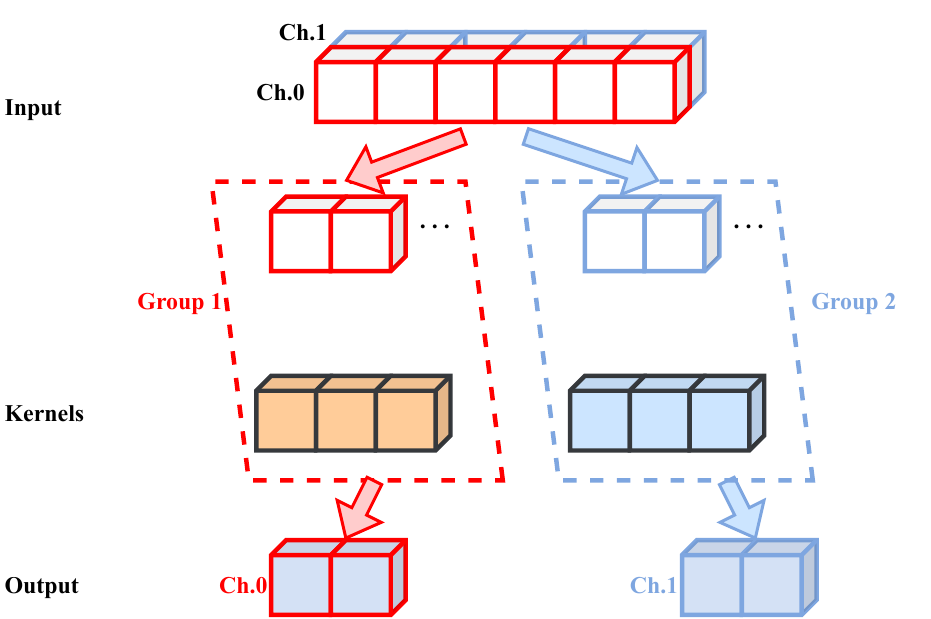}
    \caption{}
    \label{fig_multiChanConv}
\end{subfigure}
\caption{(a) Diagram of the operation of a multi-channel transposed convolutional layer with $2$ input channels, $4$ output channels, and $2$ processing groups. (b) Diagram of the operation of a multi-channel convolutional layer with $2$ input channels, $2$ output channels, and $2$ processing groups.}
\label{fig_multiChanLayer}
\vspace{-3mm} 
\end{figure}

As illustrated in Figure~\ref{fig_multiChanLayer}, both convolutional and transposed layers support the processing of multiple input and output channels. Additionally, the grouping feature of these layers can divide the input and output channels into groups, enabling the independent processing for separate groups. These features serve as the crucial basics for handling the multiple branches of polyphase filter banks.

Based on the above demonstration, we derive the neural network models that can be used for DSP operations required by filter banks, specifically, the DFT/IDFT and interpolation/decimation.

\begin{figure}[t]
    \centering
    \includegraphics[keepaspectratio=true, width=0.8\linewidth]{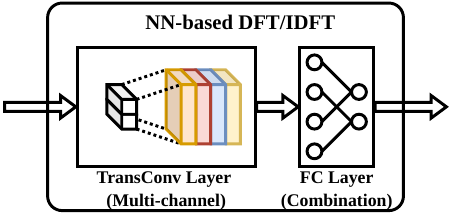}
    \caption{Diagram of the NN-based DFT/IDFT block.}
    \label{fig_nnDFT}
    \vspace{-3mm}
\end{figure}

\textbf{NN-based DFT/IDFT.} The DFT/IDFT processes share similar operations, allowing us to transform both using the same neural network model, differing only in parameter configuration. The structure of NN-based DFT/IDFT blocks is depicted in Figure~\ref{fig_nnDFT}, consisting of a multi-channel $2$D transposed convolutional layer followed by a fully connected layer.

\begin{figure}[t]
     \centering
     \begin{subfigure}[b]{0.9\linewidth}
         \centering
         \includegraphics[keepaspectratio=true,width=\linewidth]{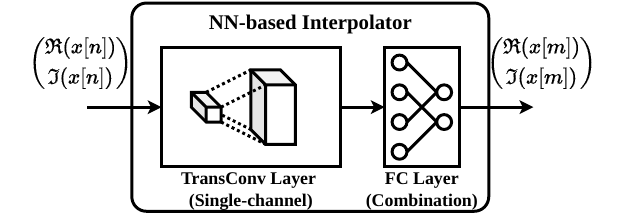}
         \caption{}
         \label{fig_nninterp}
     \end{subfigure}
     \begin{subfigure}[b]{0.9\linewidth}
         \centering
         \includegraphics[keepaspectratio=true,width=\linewidth]{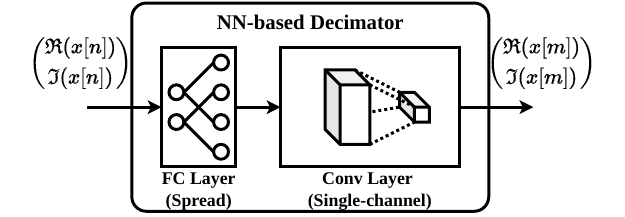}
         \caption{}
         \label{fig_nndecim}
     \end{subfigure}
     \vspace{-4mm}
     \label{fig_nnRateConverter}
     \caption{(a) Diagram of the NN-based interpolation block. (b) Diagram of the NN-based decimation block.}
     \vspace{-5mm}
\end{figure}

\textbf{NN-based Interpolation.} We utilize the transposed convolutional layer to perform both upsampling and filtering in the interpolation process. The corresponding neural network model is illustrated in Figure~\ref{fig_nninterp}.

For cases where both the input signal and the anti-imaging filter are complex-valued, we represent the complex values using their real and imaginary parts. We then apply a 2D transposed convolutional layer to process these 2D inputs. \textbf{The kernels of the transposed convolutional layer are configured to match the real and imaginary components of the anti-imaging filter, with the stride set according to the upsampling factor.} The output of the 2D transposed convolution forms the components representing the complex-valued signals. A subsequent fully-connected layer combines these components to produce the final output.

\textbf{NN-based Decimation.} Inspired by the similar mechanisms in Equation~\ref{eq_decim} and the convolutional layer, we developed an NN-based decimation block, as shown in Figure~\ref{fig_nndecim}. This block consists of a fully connected layer followed by a 2D convolutional layer. 

Similarly, the proposed design can handle cases where both the input signal and the anti-aliasing filter are complex-valued. The fully connected layer expands the input signal tensor, after which the 2D convolutional layer processes the expanded tensor to generate the decimation output. \textbf{The kernels of the 2D convolutional layer are configured based on the real and imaginary components of the anti-aliasing filter, with the stride set according to the downsampling factor.}

\begin{figure*}[ht!]
     \centering
     \begin{subfigure}[b]{0.49\textwidth}
         \centering
         \includegraphics[keepaspectratio=true,width=\linewidth]{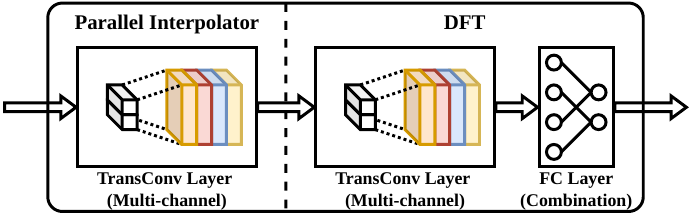}
         \caption{}
         \label{fig_nnpafb}
     \end{subfigure}
     \begin{subfigure}[b]{0.49\textwidth}
         \centering
         \includegraphics[keepaspectratio=true,width=\linewidth]{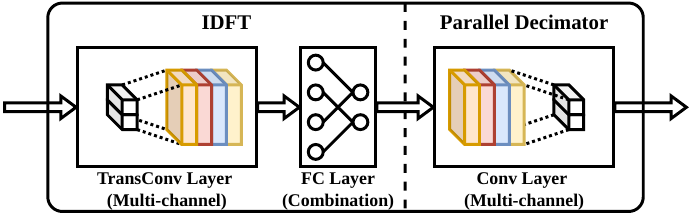}
         \caption{}
         \label{fig_nnpasb}
     \end{subfigure}
     \vspace{-2mm}
     \caption{Diagram of (a) the NN-based polyphase analysis filter bank and (b) the NN-based polyphase synthesis filter bank.}
     \label{fig_nnpfb}
     \vspace{-3mm}
\end{figure*}

\subsection{NN-based Filter Banks}
The templates for the NN-based Polyphase Analysis Filter Bank (NNPAFB) and NN-based Polyphase Synthesis Filter Bank~(NNPSFB) are shown in Figure~\ref{fig_nnpfb}. Instead of directly stacking the NN-based DSP blocks to form the filter banks, we make several simplifications to the architectures. These modifications rely on the characteristics and features of the practical filter design and neural network layers.

\textbf{Multi-channel Parallelism.} We leverage the multi-channel capabilities of neural network layers to enable parallel interpolation or decimation processes as in Section~\ref{subsec_ppfb}. More specifically, the channels for the NN-based Interpolation/Decimation networks are configured to match the number of sub-bands for PAFB and PSFB. 

\textbf{Model Simplification.} We further simplify the models by considering the characteristics of filters, as the prototype analysis and synthesis filters can be implemented with real-valued taps. Therefore, we can eliminate the kernels associated with the imaginary parts of the filters and retain only those related to the real parts. Additionally, the fully connected layers in the NN-based interpolators and decimators are also simplified as they are designed to handle complex-valued filtering processes. Reducing the number of parameters not only streamlines the model but also facilitates faster convergence during the training process, which will be discussed in the following subsections.

\begin{figure}[t]
    \centering
    \includegraphics[keepaspectratio=true, width=\linewidth]{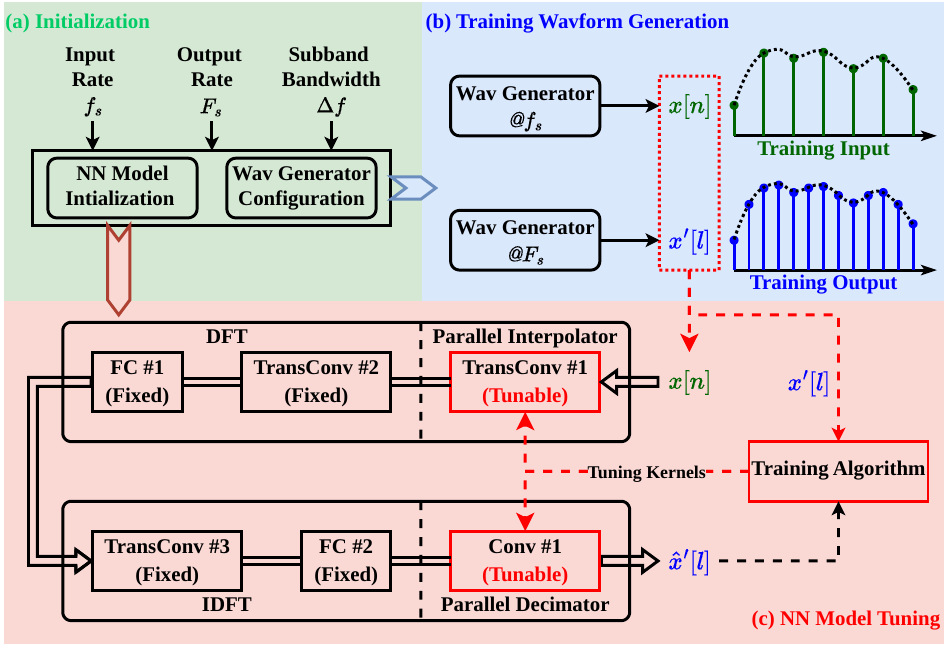}
    \caption{The training procedure for NN-based filter banks. (a) Initialization. (b) Training waveform generation. (c) NN model training.}
    \label{fig_nnpfb_training}
    \vspace{-5mm}
\end{figure}

\subsection{Filter Weight Configuration}
To ensure distortionless spectrum multiplexing, the filters used in both the analysis and synthesis filter banks require careful design. Traditionally, these filters are designed analytically, a process that demands expert knowledge in filter design. 

The analysis filter mainly serves the functionality for a narrow-band extraction, meaning that it should have a flat frequency response within the desired passband. Due to the finite impulse response, the frequency response has a transition band, containing spectrum components from adjacent subbands, which imposes challenges to the synthesis filter. The passband of the synthesis filter should be large enough to cover the passband as well as the transition band of the synthesis filter. Meanwhile, it's supposed to have a steep transition to filter out the imaging components caused by upsampling. 

Generally, given the bandwidth of the subband signals and the sampling rate of the input stream, the analysis filter is configured based on windowed \textit{sinc} filters with closed-form expressions. The synthesis filters are formulated through frequency-domain approximation, which requires solving complicated optimization problems~\cite{crochiereMultirateDigitalSignal1983,vaidyanathanMultirateDigitalFilters1990}. Although there are some filter design toolboxes~\cite{MatlabSP, scipy}, the key parameters, such as the cutoff frequency, the roll-off region, etc., still require trial-and-error, increasing the design complexity.

The adoption of neural network models for filter banks introduces learning capabilities, allowing the filters to be configured through training. Inspired by pilot-aided channel estimation in wireless communication~\cite{goldsmith2005wireless}, we propose using analytically manipulated signals as training sets to learn the filter taps.

The training workflow involves three key steps: initialization, generation of training signals, and tuning of the parameters.

\textbf{Initialization.} Based on the relationship between the input and output signals and the bandwidth of the desired decomposed subband, we initialize the cascaded NN models for the filter banks. 

The interpretable structure of the proposed NN-based filter banks allows for a significant reduction in the number of trainable parameters. According to our design, the NN-based DFT/IDFT blocks within the filter banks can be configured with predefined parameters based on the DFT/IDFT expressions. Additionally, the fully connected layers for complex-valued representation have fixed weights. As a result, the only trainable parameters are the kernels in the transposed convolutional layer (\textit{TransConv \#1}) in the NNPAFB and the convolutional layer (\textit{Conv \#1}) in the NNPSFB. Meanwhile, we reduce the size of the kernels as we only consider the real-valued filter taps. 

Moreover, we incorporate symmetry from the design principles of linear-phase filters to further constrain the trainable parameters. For example, for a filter of length $N+1$ (with $N$ being even), we make the first $N/2+1$ parameters trainable while the remaining $N/2$ parameters are mirrored symmetrically to the first half.

Additionally, we utilize a model-driven initialization technique for faster tuning. The trainable kernels are initialized based on the \textit{sinc} functions, which are commonly used as prototypes in filter design. More specifically, the \textit{sinc} filters are decomposed following the principles in Section~\ref{sec_fbmux}, and the kernels are initialized based on these polyphase components. 

\textbf{Training waveform generation.} Given the configuration of the input-output relation, we use waveform generators to produce training sets based on analytical formulas. In order to make the filter banks more general, we apply the waveform generators to generate signals from various schemes. The waveform generators are designed to operate at different sampling rates, as illustrated in Figure~\ref{fig_nnpfb_training}. A straightforward approach is to apply modulators to the same batch of symbols. The modulators have the same characteristics but work at different sampling rates. As a result, we obtain signal samples, $x[n]$ and $x'[l]$, that exhibit similar waveforms but differ in their sampling rates.

\textbf{Tuning NN-based filter banks.} During the tuning process, we feed the low-sample-rate signals, $x[n]$, into the cascaded NN-based filter banks for a basic multiplexing task. The goal is to minimize the error between the output from the NN models, $\hat{x}'[l]$, and the target training signals, $x'[l]$.

%% file: Sections/6_Implementation.tex
\section{Implementation}\label{sec_impl}
In this work, we use PyTorch~\cite{pytorch} as the implementation framework. The code is available in GitHub repository\footnote{https://github.com/Repo4Sub/Sensys2026}. The neural layers utilized in our design include \textit{ConvTranspose2d}, \textit{Conv2d}, and \textit{Linear} layers. In practice, we design some customized layers to achieve the parameter simplification for training purposes and then embed the trained kernels into standard built-in neural layers for inference or evaluation.

We also deploy the spectrum multiplexer as a plug-in module within the software-defined radio~(SDR) system workflow. For testing, we select two typical host devices: a high-performance desktop and an embedded single-board computer (SBC) with ARM-based processors. The desktop is equipped with an Intel i7-13700KF CPU and an Nvidia RTX 4070 Ti GPU, while the SBC is the Nvidia Jetson Orin Kit~\cite{jetsonorin}, which also features a built-in Nvidia GPU. In both cases, we primarily utilize the GPU to provide hardware acceleration.

%% file: Sections/7_Evaluation.tex
\section{Evaluation}\label{sec_eval}
We evaluate the NN-based spectrum multiplexer in two key aspects: waveform quality and efficiency.

\textbf{Performance Metrics:} To assess waveform quality, in addition to direct waveform visualization, we focus on the receivers' ability to correctly receive the multiplexed signals. The evaluation metrics include Normalized Mean Square Error (NMSE), Bit Error Rate (BER), and Packet Reception Ratio (PRR). For efficiency, we primarily evaluate the execution time.

\textbf{Baseline Methods:} We compare the proposed NNPFB-based spectrum multiplexer with both the direct interpolation-based multiplexing technique and the DFT-based method. All these baseline methods are implemented in Python using SciPy~\cite{scipy}, a widely recognized scientific computing library. 

\subsection{Comparison with Plain NN Models}
We begin by conducting comparisons with plain neural network models. Since no existing models are specifically designed for the spectrum multiplexing task, we adopt two common NN models used in audio super-resolution and image processing: U-Net-based~\cite{ronneberger2015u, kuleshov2017audio} and ResNet-based~\cite{he2016deep} architectures. Specifically, we modify these models to emulate the interpolation process for spectrum multiplexing.

All models are designed to perform $2$x and $4$x interpolation processes. The training signals include pulse-shaped QPSK signals, ZigBee signals, and BLE signals, all generated through standard baseband modulation processing, as discussed before. We then test these models by interpolating randomly generated QPSK, ZigBee, and BLE signals in the validation sets. The settings for the training process are detailed in Table~\ref{tab_training}. Notably, the number of trainable parameters in the U-Net-based and ResNet-based models is significantly larger than in the proposed design. 

\begin{table}[ht]
\begin{tabular}{|c|ccl|}
\hline
Model                & \multicolumn{1}{c|}{UNet-based} & \multicolumn{1}{c|}{ResNet-based} & NNPFB \\ \hline
Signal Type & \multicolumn{3}{c|}{Shaped QPSK, ZigBee, BLE}                               \\ \hline
Set Size    & \multicolumn{3}{c|}{90 QPSK, 45 ZigBee, 45 BLE}                                              \\ \hline
\# of Parameters           & \multicolumn{1}{c|}{153828}     & \multicolumn{1}{c|}{144546}       & \multicolumn{1}{c|}{127}   \\ \hline
\# of Epochs               & \multicolumn{3}{c|}{200}                                                    \\ \hline
\end{tabular}
\caption{The settings of the training processes for the U-Net-based and ResNet-based model and the proposed NN-based Polyphase Filter Bank~(NNPFB).}
\label{tab_training}
\vspace{-5mm}
\end{table}

\begin{figure}[ht]
     \centering
     \begin{subfigure}[b]{0.9\linewidth}
         \centering
         \includegraphics[keepaspectratio=true,width=\linewidth]{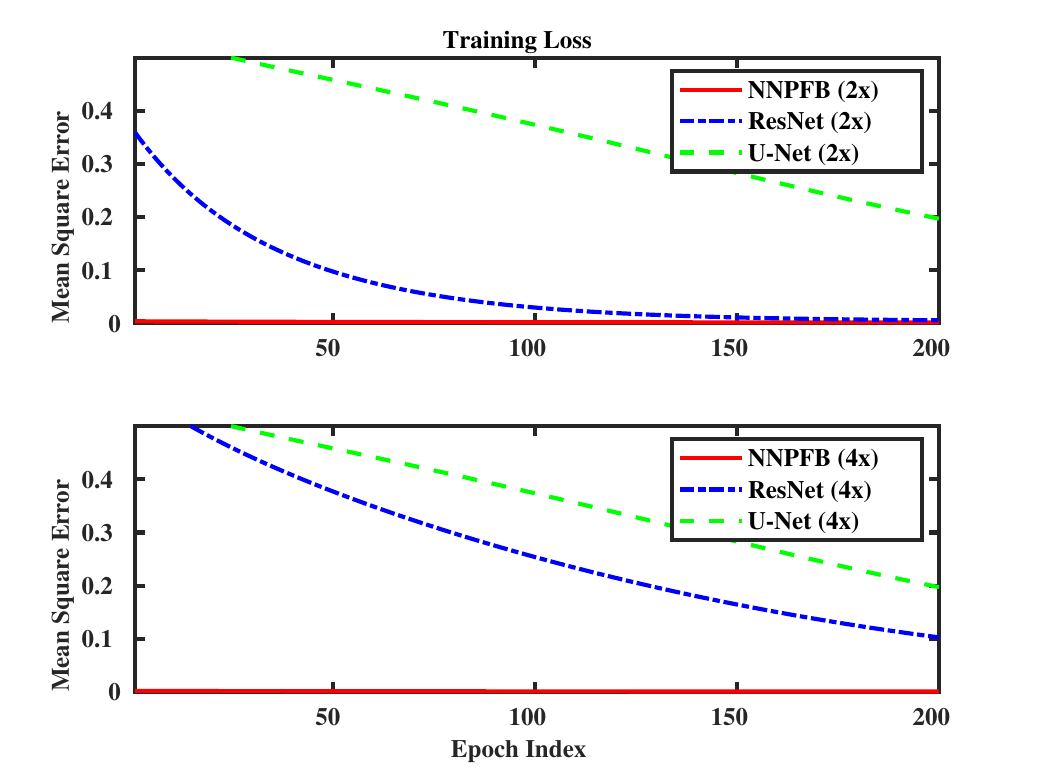}
         \caption{}
         \label{fig_trainingloss}
     \end{subfigure}
     \begin{subfigure}[b]{0.9\linewidth}
         \centering
         \includegraphics[keepaspectratio=true,width=\linewidth]{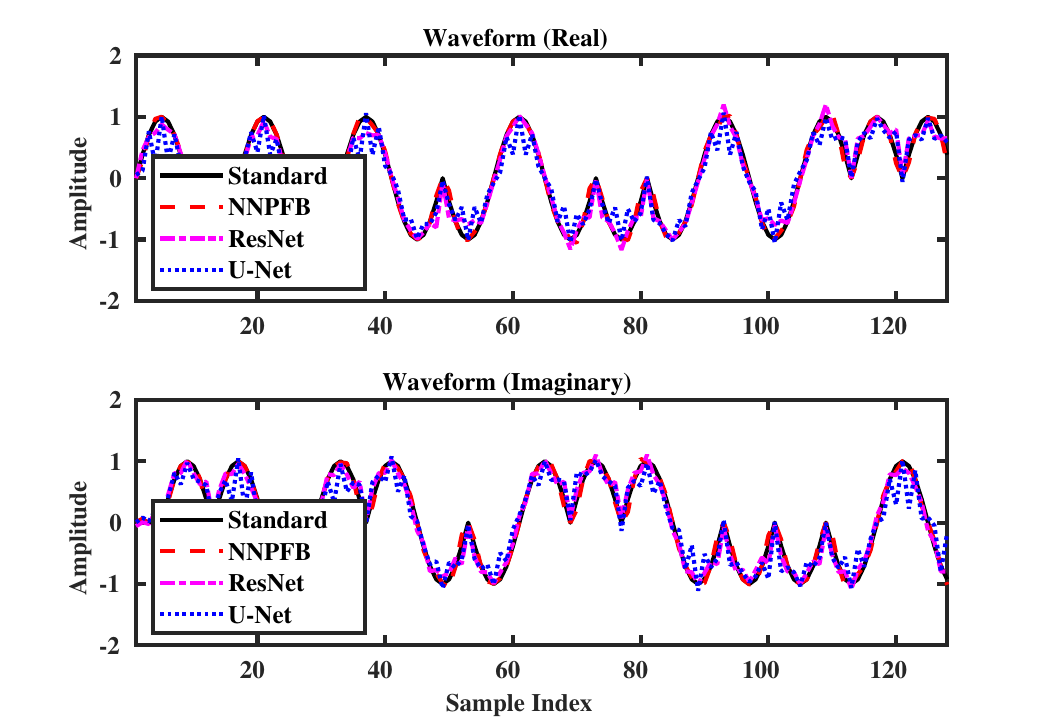}
         \caption{}
         \label{fig_trainingWav}
     \end{subfigure}
     \vspace{-3mm}
     \caption{(a) Training loss of the plain U-Net-based, ResNet-based and NNPFB-based filter bank models. (b) The $2$x interpolated signals for ZigBee baseband signals from three models.}
     \vspace{-3mm}
\end{figure}

The low-sampling-rate signals are fed into the models, and all models are trained to minimize the Mean Square Error (MSE) between their output signals and the high-sampling-rate training signals. The MSE loss during training and tuning is plotted in Figure~\ref{fig_trainingloss}. The initial loss values for the proposed NNPFB are significantly lower than those of the other two models. This is because the parameters in our design have a direct relationship with DSP models, allowing us to initialize them with model-driven filter weights.

After the training process, the models are applied to interpolate the validation signals, which include general QPSK, ZigBee, and BLE signals. A direct comparison of the interpolated waveforms is shown in Figure~\ref{fig_trainingWav}. It is clear that our NNPFB method closely matches the standard reference waveform. We also compute the Normalized Mean Square Error (NMSE) to quantify the distortion level, as defined in Equation~\ref{eq_nmse}. The results, listed in Table~\ref{tab_nmse}, show that the proposed NNPFB-based method achieves the lowest distortion level compared to the other two models.

Thus, while plain models may not demand extensive expert knowledge, our specific NNPFB design, rooted in DSP models, significantly outperforms them in terms of waveform quality.

\begin{equation}
    \text{NMSE}(x,y) = 10\log(\frac{\sum|\hat{x}'[l]-x'[l]|^2}{\sum|x'[l]|^2})
    \label{eq_nmse}
\end{equation}

\begin{table}[ht]
\begin{tabular}{|cccc|}
\hline
\multicolumn{4}{|c|}{$2$x Interpolation}                                                                                    \\ \hline
\multicolumn{1}{|c|}{Methods}      & \multicolumn{1}{c|}{U-Net-based} & \multicolumn{1}{c|}{ResNet-based} & NNPFB           \\ \hline
\multicolumn{1}{|c|}{General QPSK} & \multicolumn{1}{c|}{-19.69}      & \multicolumn{1}{c|}{-21.89}       & \textbf{-33.01} \\ \hline
\multicolumn{1}{|c|}{BLE (GMSK)}   & \multicolumn{1}{c|}{-22.62}      & \multicolumn{1}{c|}{-25.09}       & \textbf{-35.89} \\ \hline
\multicolumn{1}{|c|}{ZigBee}       & \multicolumn{1}{c|}{-17.18}      & \multicolumn{1}{c|}{-20.43}       & \textbf{-33.36} \\ \hline
\multicolumn{4}{|c|}{$4$x Interpolation}                                                                                    \\ \hline
\multicolumn{1}{|c|}{General QPSK} & \multicolumn{1}{c|}{-9.16}       & \multicolumn{1}{c|}{-9.15}        & \textbf{-39.49} \\ \hline
\multicolumn{1}{|c|}{BLE (GMSK)}   & \multicolumn{1}{c|}{-11.48}      & \multicolumn{1}{c|}{-10.84}       & \textbf{-38.27} \\ \hline
\multicolumn{1}{|c|}{ZigBee}       & \multicolumn{1}{c|}{-8.26}       & \multicolumn{1}{c|}{-10.00}       & \textbf{-29.87} \\ \hline
\end{tabular}
\caption{The NMSE values of different methods to interpolate different types of signals. Lower NMSE means less distortion.}
\label{tab_nmse}
\vspace{-3mm}
\end{table}

\subsection{Interpretable Trained NNs}
We also validate the interpretability of the parameters within the NN-based filter banks. The trained kernels are considered the polyphase components of the analysis or synthesis filters, so that we can recover the \textit{trained analysis or synthesis filters} from trained kernels. 

Here, we discuss the training settings for the NN-based filter banks in more detail. Similar to the previous, we consider the $2$x interpolation task. Without loss of generality, the NNPAFB is supposed to decompose the input signals of $8$MHz sample rate into oversampled subband signals with an interval of $0.5$MHz and a sample rate of $2\times0.5=1$MHz. The NNPSFB will combine these sub-band signals to form the output with a sample rate of $16$MHz. Following the discussion in Section~\ref{sec_fbmux}, the kernels in NNPAFB are fixed based on a $sinc$ filter multiplied by the \textit{Kaiser} window, which intends to have a normalized bandwidth of $\pi/8$. Meanwhile, the kernels in NNPSFB are initialized with a truncated \textit{sinc} filter with a normalized bandwidth of $\pi/8$ and tuned with the training sets. 

To verify that our trained kernels in NNPFB can benefit from the model-driven initialization, we also trained the NNPSFB with another different type of initialization value based on the trivial normalization method. We utilized the built-in initialization method by randomly generating from a normal distribution.

\begin{figure}[ht]
     \centering
     \includegraphics[keepaspectratio=true,width=\linewidth]{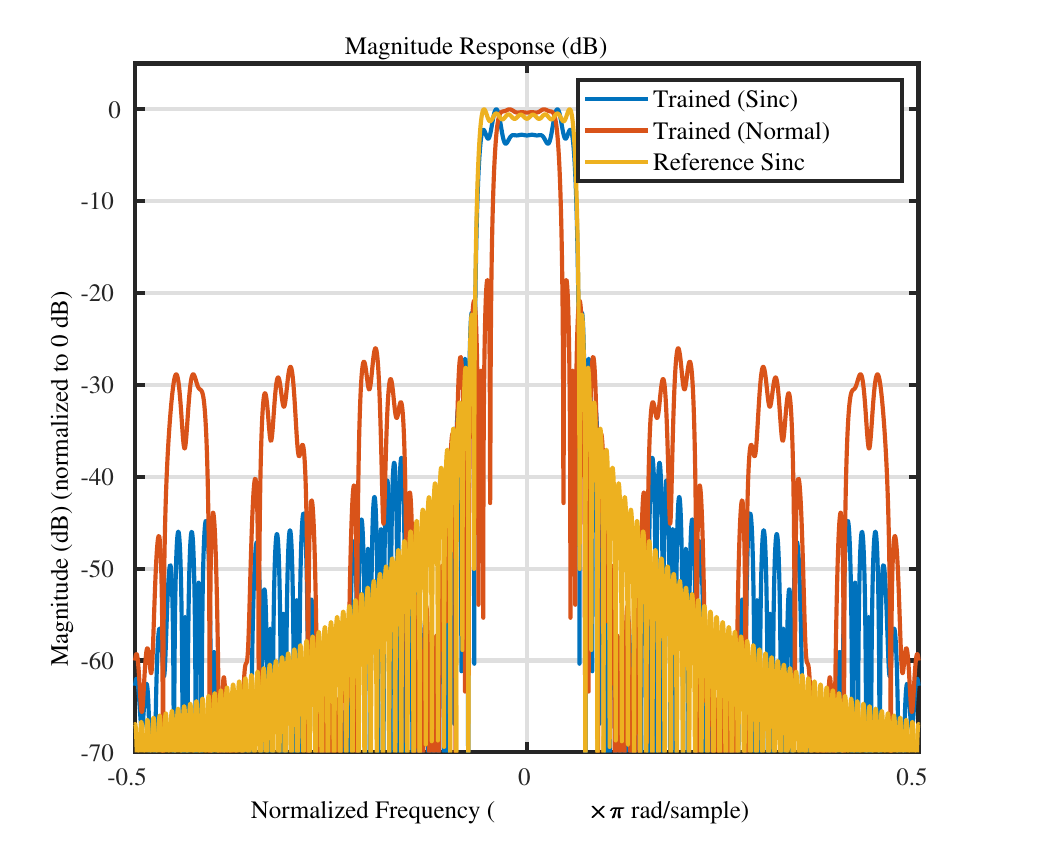}
     \caption{The trained synthesis filters with different initialization.}
     \label{fig_weight}
     \vspace{-3mm}
\end{figure}

We plot the amplitude-frequency response of the reference \textit{sinc} filter and the trained synthesis filters in Figure~\ref{fig_weight}. As depicted in the figure, both trained filters have a similar pattern, of which the amplitude-frequency response levels around $\omega=\pm k\pi/8 $ are low, meaning that the NNPSFBs are trained to minimize the image components caused by upsampling. At the first image region around $\omega=\pm \pi/8 $, which is the closest to the passband, both trained filters reach lower levels than the reference \textit{sinc} filter. Meanwhile, the trained filter with model-driven \textit{sinc}-based initialization has a larger overall attenuation than that with normal-distribution initialization, resulting in less distortion.

These results demonstrate that the proposed NNPFB can be trained to derive the prototype filters required for spectrum multiplexing tasks, which not only significantly reduces the design complexity of conventional handcrafted designs but also shows the potential to suppress the distortion.

\subsection{Waveform Quality}
Next, we verify the quality of the generated waveform in a simulation environment. For simplicity, we simulate a single Wi-Fi stream by increasing the sampling rate and shifting the central frequency of the baseband spectrum. The simulation transmission configurations are discussed below.

Suppose we intend to transmit IEEE 802.11n (Wi-Fi-4) packets on Channel $11$, with a central frequency of $f_c = 2462$ MHz and a bandwidth of $BW = 20$ MHz. The IEEE 802.11n packets are generated with a sample rate of $f_s = 20$MHz by using Matlab~\cite{MatlabWlan}. We assume that the radio front-end is configured to operate at a central frequency of $f_c' = 2460$ MHz and a sample rate of $f_s' = 40$MHz. As a result, the baseband signals must be interpolated to match the higher sample rate and modulated by the frequency offset.

We use three methods to accomplish this task: two baseline conventional DSP methods—the direct interpolation-and-modulation approach and the DFT-based approach—along with our NNPFB method. The generated signals are transmitted through a simulated AWGN channel. On the receiver side, the Wi-Fi baseband signals are processed by a standard Wi-Fi receiver working at $f_c' = 2460$~\cite{MatlabSP}.

\begin{figure}[ht]
     \centering
     \begin{subfigure}[b]{0.9\linewidth}
         \centering
         \includegraphics[keepaspectratio=true,width=\linewidth]{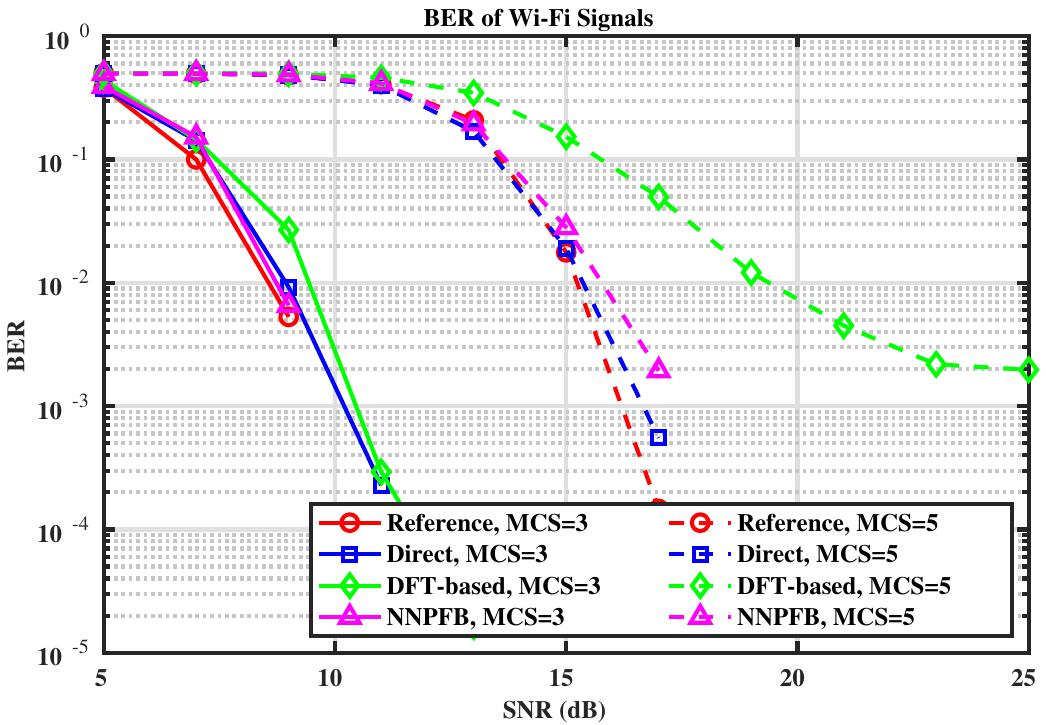}
         \caption{}
         \label{fig_ber_wifi}
     \end{subfigure}
     \begin{subfigure}[b]{0.9\linewidth}
         \centering
         \includegraphics[keepaspectratio=true,width=\linewidth]{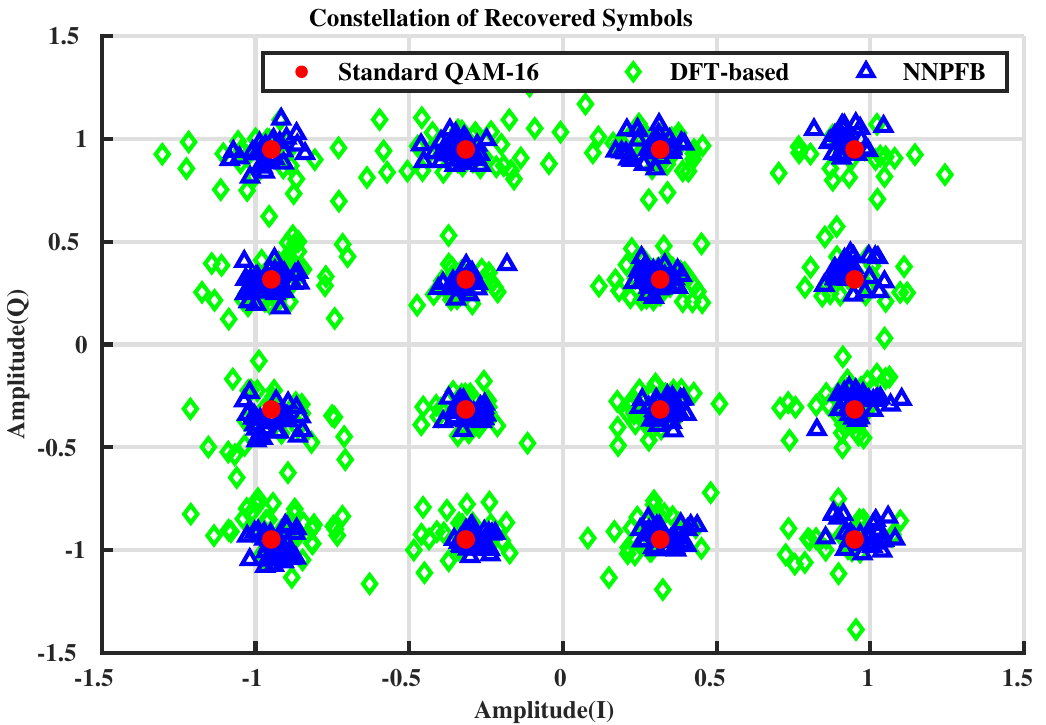}
         \caption{}
         \label{fig_const_wifi}
     \end{subfigure}
     \vspace{-3mm}
     \caption{(a) BER curves of signals generated by different approaches. (b) The recovered symbols of the DFT-based method and our NNPFB method (SNR$=20$dB). }
     \vspace{-5mm}
\end{figure}

We simulate various settings of Wi-Fi packets and plot the BER curves in Figure~\ref{fig_ber_wifi}. Both the direct interpolation approach and the proposed NNPFB design achieve performance comparable to the standard reference signals, while the DFT-based approach shows lower accuracy, largely due to inherent spectral leakage. The NNPFB method has up to $10$dB SNR gain compared with the DFT-based approach. Additionally, we validate the recovered symbols by plotting their constellations in Figure~\ref{fig_const_wifi}. The NNPFB method introduces minimal distortion, resulting in denser constellation clusters at the receiver. In contrast, the DFT-based method exhibits severe distortion during transmission, leading to more chaotic symbol distributions, even at a high SNR (SNR $=25$ dB), which should be sufficient for MCS $=3$ (requiring SNR $\geq12.5$ dB). 

\subsection{Efficiency}
The following evaluations focus on efficiency, particularly in the practical application of handling multiple input streams. The configurations related to the methods are outlined below.

We multiplex three ZigBee streams onto three adjacent channels using conventional methods and our NNPFB-based method. The ZigBee signals are generated with a sample rate of $4$MHz, and the interval between ZigBee channels is $5$MHz. So, we consider a target wide band of $16$MHz, and three streams have the frequency offsets of $\{-5, 0, 5\}$MHz. 

In the direct approach, an interpolator with an upsampling factor of $16/4=4$ is applied, along with an anti-imaging filter with a cutoff frequency at a normalized value of $0.25\pi$ and $128$ taps. The frequency offset sequences are generated as described in Figure~\ref{fig_specmux_time}.

The DFT-based approach performs an $8$-point DFT on each ZigBee stream, obtaining frequency-domain components with a resolution of $4/8 = 0.5$ MHz. Based on the frequency offsets and resolution, the corresponding frequency domain shifts are $\{-10, 0, 10\}$ bins. The multiplexed signal is generated by taking the IDFT on 32 frequency bins ($16/0.5 = 32$). Of these, $3 \times 8 = 24$ bins are allocated to the frequency-domain components of the three ZigBee streams, while the remaining bins are set to zero.

For the proposed NNPFB-based method, we decompose the ZigBee streams into $8$ sub-bands using NNPAFB and then combine $32$ sub-bands at the NNPSFB. The neural network model weights are configured based on prior training, and the mapping of subband signals to the synthesis filter bank inputs is derived based on the frequency shifts and the intervals between subbands.

\begin{figure}[ht]
     \centering
     \begin{subfigure}[b]{0.9\linewidth}
         \centering
         \includegraphics[keepaspectratio=true,width=\linewidth]{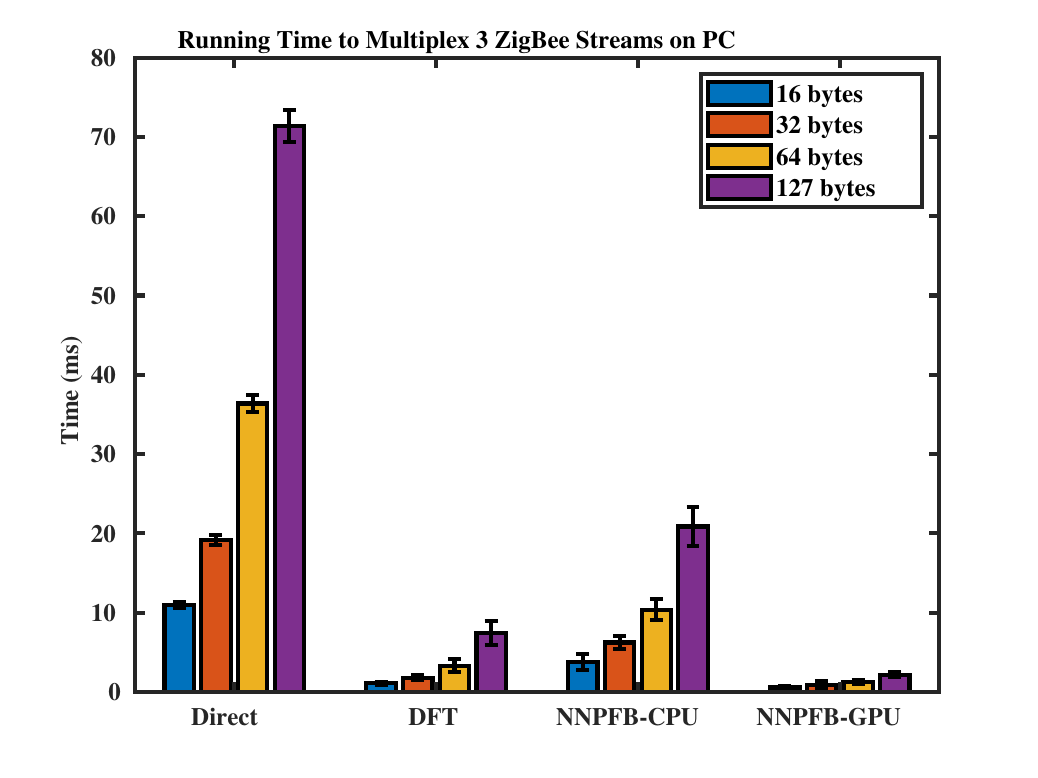}
         \caption{}
         \label{fig_time_pc}
     \end{subfigure}
     \begin{subfigure}[b]{0.9\linewidth}
         \centering
         \includegraphics[keepaspectratio=true,width=\linewidth]{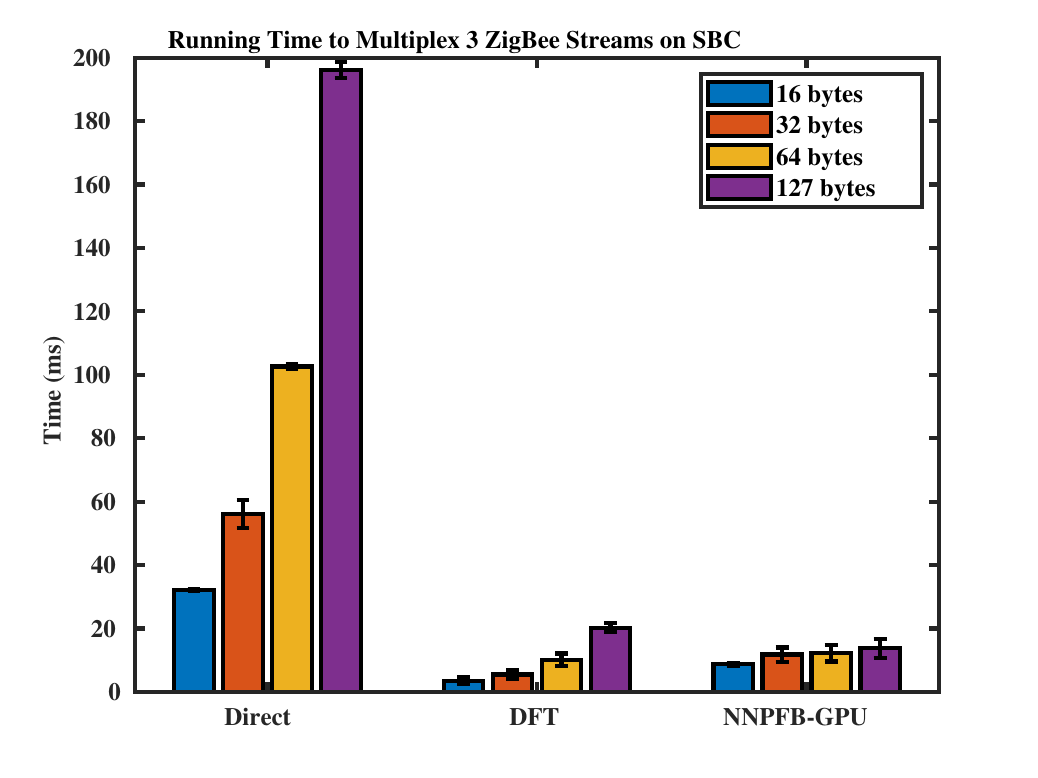}
         \caption{}
         \label{fig_time_sbc}
     \end{subfigure}
     \vspace{-3mm}
     \caption{(a) Running time on desktop PC. (b) Running time on a single board computer.}
     \vspace{-3mm}
\end{figure}

All methods process $10$ ZigBee packets per stream with varying message sizes. We measure the running time required to complete the spectrum multiplexing tasks on both the desktop PC and the SBC, with the NNPFB method configured to utilize GPU acceleration. The results are displayed in Figure~\ref{fig_time_pc} and Figure~\ref{fig_time_sbc}.

As shown in Figure~\ref{fig_time_pc}, when the message length doubles, the running time for all methods increases nearly exponentially. The direct approach has the worst performance on both platforms, as the modulation process requires element-wise complex-valued multiplication. For ZigBee packets with the maximum message length, the direct approach takes up to $71$ milliseconds to complete. The DFT-based method and our NNPFB method are more efficient than the direct approach, as they simplify the heavy modulation processes by indexing and mapping the frequency-domain components or sub-band signals accordingly. The running time of the NNPFB method without acceleration is slightly longer than that of the DFT-based method due to the extra filtering processes. However, when configured with GPU acceleration, the NNPFB method performs the multiplexing tasks in just $2$ milliseconds—nearly $10$ times faster than without acceleration, $3$ times faster than the DFT-based method, and $35$ times faster than the direct approach.

Efficiency is also influenced by the computational compatibility of the platforms. As shown in Figure~\ref{fig_time_sbc}, the running time for the conventional direct and DFT-based methods on the embedded systems increases significantly, highlighting the need for further optimization. In contrast, the NNPFB method can be easily configured to run with acceleration, as neural network frameworks offer native support for hardware acceleration. With GPU acceleration, the NNPFB method performs up to $2$ times faster than the DFT-based method and 15 times faster than the direct approach.

It is also important to note that although the DFT-based method demonstrates comparable efficiency, the previous results show that its distortion can lead to communication performance degradation. This undermines the primary goal of the spectrum multiplexer.

\subsection{Field Experiments}
We also conducted field experiments with the NNPFB-based spectrum multiplexer. The NNPFB-based spectrum multiplexers are integrated into the SDR workflow. For these experiments, we used the USRP X310 ~\cite{USRPx310} SDR front-end, which operates in the typical $2.4$GHz ISM band for common IoT schemes and supports a sample rate of up to $200$MHz. We explored two application scenarios: multi-user communication for a homogeneous IoT scheme and signal combination for heterogeneous IoT schemes.

\textbf{Multiple ZigBee Transmission} We extend the previous evaluation on three ZigBee streams into real-world experiments. The settings related to the field experiments are listed below (Table~\ref{tab_zig_test}). As for the NNPFB design, we follow the same settings as before. It's also worth pointing out that we can apply a single analysis filter bank to handle three streams because all the streams share the same sample rate of $4$MHz.

\begin{table}[ht]
\begin{tabular}{|c|ccc|c|}
\hline
                  & \multicolumn{3}{c|}{ZigBee Baseband}                         & USRP Front-end \\ \hline
Streams           & \multicolumn{1}{c|}{Z-1}  & \multicolumn{1}{c|}{Z-2}  & Z-3  & FE-1           \\ \hline
Frequency (MHz)   & \multicolumn{1}{c|}{2470} & \multicolumn{1}{c|}{2475} & 2480 & 2475           \\ \hline
Sample Rate (MHz) & \multicolumn{3}{c|}{4}                                       & 16             \\ \hline
\end{tabular}
\caption{Settings for multiple ZigBee streams via NNPFB-based spectrum multiplexer. The USRP front-end is configured based on the requirements for each stream.}
\label{tab_zig_test}
\vspace{-7mm}
\end{table}

To validate the spectrum multiplexing, we placed three ZigBee receivers (TI CC2650 Kit~\cite{TIpad}) in a standard room environment. The multiplexed signals were transmitted over the air using the USRP front-end, and we measured the Packet Reception Ratio~(PRR) at each receiver by counting the error-free packets. As a baseline, we used the USRP to transmit only one ZigBee stream at a time and measured the PRR at the corresponding receiver. The PRR results, shown in Table~\ref{tab_zigbee_prr}, indicate that the multiplexed signals achieved a similar PRR to the single-stream transmissions.

\begin{table}[ht]
\begin{tabular}{|c|c|c|c|}
\hline
Receiver & Z-1     & Z-2     & Z-3      \\ \hline
Distance & 3m, LoS & 6m, LoS & 5m, NLoS \\ \hline
Baseline & 97\%    & 95\%    & 83\%     \\ \hline
NNPFB    & 98\%    & 95\%    & 81\%     \\ \hline
\end{tabular}
\caption{The PRRs of multiplexed three ZigBee streams and single stream transmission.}
\label{tab_zigbee_prr}
\vspace{-7mm}
\end{table}



\textbf{Wi-Fi and ZigBee Combination} We also used the NNPFB-based spectrum multiplexer to handle signals from different IoT schemes with varying sample rates. In this case, we combined a $20$ MHz Wi-Fi signal and a $4$ MHz ZigBee signal for simultaneous transmission through a single USRP front end. The settings for this experiment are listed in Table~\ref{tab_zigwifi_test}. As in the previous simulation, the USRP front-end was configured to operate at a central frequency of $2460$ MHz with a sample rate of $40$ MHz.

\begin{table}[ht]
\begin{tabular}{|c|c|c|c|}
\hline
                  & ZigBee & Wi-Fi & USRP Front-end \\ \hline
Frequency (MHz)   & 2475   & 2462  & 2460           \\ \hline
Sample Rate (MHz) & 4      & 20    & 40             \\ \hline
\end{tabular}
\caption{Settings to combine ZigBee and Wi-Fi signals via NNPFB-based spectrum multiplexer..}
\label{tab_zigwifi_test}
\vspace{-7mm}
\end{table}

\begin{figure}[ht]
    \centering
    \includegraphics[keepaspectratio=true, width=\linewidth]{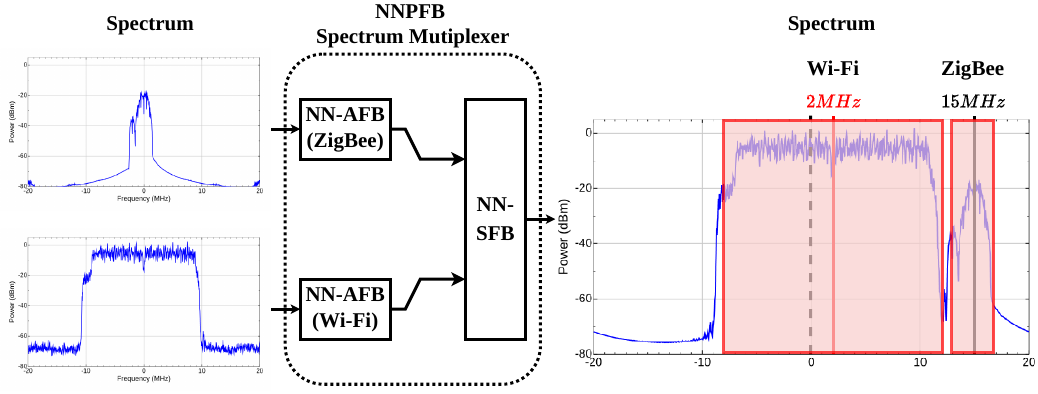}
    \caption{Diagram of Wi-Fi and ZigBee signal combination via NNPFB spectrum multiplexer.}
    \label{fig_zigwifi}
    \vspace{-5mm}
\end{figure}

The workflow is illustrated in Figure~\ref{fig_zigwifi}. Since the two baseband signals have different sample rates, we design two distinct analysis filter banks. The decomposed subband signals are mapped to the input of a shared synthesis filter bank, which generates the final multiplexed signals.

For validation, we visualized the spectrum of the Wi-Fi, ZigBee, and combined signals. As shown in the figure, the NNPFB spectrum multiplexer successfully shifts the Wi-Fi and ZigBee signals to their designated portions of the spectrum. We also measured the PRRs for both streams in an indoor LoS environment: the Wi-Fi receiver achieved a PRR of $95\%$, while the ZigBee receiver achieved $96\%$.

%% file: Sections/8_Discussion.tex
\section{Related Work}~\label{sec_related}
\textbf{Spectrum Multiplexer} The research on radio virtualization~\cite{sachsVirtualRadioFramework2008,santosVirtualRadiosReal2020, priceTransceiversResourceScheduling2020} proposes to multiplex signals on the spectrum. Among them, \cite{tanEnableFlexibleSpectrum2012} is a pioneer work by applying the DFT-based methods for spectrum multiplexing, followed by \cite{kistSDRVirtualizationFuture2018,defigueiredoBasebandWirelessSpectrum2020,kistAIRTIMEEndEndVirtualization2022}. \cite{liuEnablingVirtualRadio2020} applies the direct interpolation and modulation for spectrum multiplexing. Our methodology differs from these works in that we apply the filter banks for spectrum multiplexing.  

\textbf{Polyphase Filter Banks} Polyphase filter banks are well-studied in the domain of signal processing. \cite{harrisDigitalReceiversTransmitters2003,harrisCascadeNonmaximallyDecimated2017,harrisWideband160channelPolyphase2011} adopts the polyphase filter banks in communication systems for sample rate conversion. \cite{smithHighThroughputOversampledPolyphase2021} focus on the efficient implementation of filter banks. However, these works mainly rely on the structure proposed in~\cite{harrisDigitalReceiversTransmitters2003}, which still requires extra modulation for analysis and synthesis procedure. In comparison, we adopt a different structure first introduced in ~\cite{crochiereMultirateDigitalSignal1983}, which is more friendly for NN-based design.

\textbf{NN as DSP operations} Although neural networks are widely adopted for signal processing tasks, there are only a few works on using neural network layers for low-level DSP operations. \cite{engel2020ddsp, hoydis2022sionna} implement customized neural network layers as DSP operations for learning purposes. Recent research~\cite{wang2024nn,wang2024nnctc} tries to interpret the neural layers with DSP operations. Our NNPFB further enhances such interpretability with more DSP operations and more features for advanced DSP operations, such as polyphase filtering.

\section{Discussion}~\label{sec_disscussion}
The proposed NNPFB has broader potential and can be further applied in various other domains.

\textbf{Adaptive Configuration}: Currently, we manually configure the NNPFB based on empirical signal processing knowledge, leveraging its inherent connection to DSP models. In the future, we aim to develop a framework for adaptive configuration, allowing the NNPFB to adjust automatically based on the specific requirements of the signal streams.

\textbf{NNPFB for Wide-Band Receivers}: NNPFB can also be utilized on the receiver side to split narrow-band signals from a wide spectrum. The analysis filter bank decomposes the wide-band signal into subbands, and the synthesis filter banks selectively combine the subband signals for different streams.

\textbf{NNPFB for Intelligent Spectrum Management}: NNPFB provides an end-to-end trainable architecture for signal decomposition and combination. This makes it well-suited for integration with modern AI/ML models to create intelligent spectrum management systems that are capable of extracting signals of interest or performing intermediate processing to suppress interference.

\section{Conclusion}
In this work, we present NNPFB, the neural network-based polyphase filter banks, and apply them to build an efficient and distortionless spectrum multiplexer. The primary contribution of this research is bridging the gap between conventional polyphase filter bank design and modern neural network models through interpretable paradigms. We leverage the learning capabilities of neural networks and propose a training framework to simplify the design of filter banks for spectrum multiplexing. Additionally, we implement the NNPFB-based spectrum multiplexer and conduct extensive experiments to evaluate its performance. The results demonstrate that the NNPFB-based multiplexer achieves an NMSE level of $-39$ dB. Both simulated and real-world transmissions show that the multiplexed signals perform comparably to standard signals, with up to a $10$dB SNR gain over conventional DFT-based multiplexers. Moreover, the NNPFB-based multiplexer achieves up to $35$ times better efficiency across different platforms.